\documentclass[aps,reprint,twocolumn,superscriptaddress,10pt]{revtex4-2}
\usepackage{graphicx}
\usepackage{bm}
\usepackage{amsfonts}
\usepackage{amssymb}
\usepackage{gensymb}
\usepackage{bbold}
\usepackage{tabularx}
\usepackage{multirow}

\graphicspath{{./}{../figs/}{../psd/}}

\newcommand{\req}[1]{Eq.~(\ref{#1})}

\newcommand{\rfig}[1]{Fig.~\ref{#1}}

\newcommand{\tc}{T_\mathrm{C}}

\newcommand{\cri}[1]{\text{CrI}_3}
\newcommand{\cgt}[1]{\text{Cr}_2\text{Ge}_2\text{Te}_6}

\begin{document}

\title{Self-consistently renormailzed spin-wave theory of layered ferromagnets on honeycomb lattice}

\author{Vagharsh Mkhitaryan}
\affiliation{Ames Laboratory, U.S.~Department of Energy, Ames, Iowa 50011}

\author{Liqin Ke}
\affiliation{Ames Laboratory, U.S.~Department of Energy, Ames, Iowa 50011}

\date{\today}

\begin{abstract}
We develop a self-consistently renormalized spin-wave theory, within a mean-field approximation, for the two-dimensional Heisenberg ferromagnet with perpendicular easy-axis anisotropy on the honeycomb lattice, as well as its few-layer and bulk extensions.
In this method, the magnetization dependence on temperature is found as the solution of the self-consistency equation.
Furthermore, we account for the physical difference of surface and bulk layers by treating the layers as separate sublattices.
Thus, the method can be readily generalized to study various magnetic phenomena in a broad range of systems, including those comprising magnetically inequivalent sublattices.
Using our theory, we calculate the temperature-dependent magnetization for two chromium-based layered van der Waals insulating magnets, Cr$_2$Ge$_2$Te$_6$ and CrI$_3$, employing various sets of Heisenberg exchange and single-ion anisotropy values reported for these materials in the existing literature.
As expected, we observe a strong dimensionality effect where the ordering temperature is reduced and its sensitivity on the anisotropy is enhanced with the decrease of dimensionality.
\end{abstract}

\maketitle

\section{Introduction}

Two-dimensional (2D) materials are of great current interest for next-generation devices due to their unique magnetic, electronic, and optoelectronic properties~\cite{ geim2013n, song2018s, wu2020ss, soriano2020nl}.
Notably, 2D magnetic materials are important for spintronics applications, as these materials and their integration in various van der Waals heterostructures open up new prospectives for the observation of novel exotic effects absent in 3D~\cite{geim2013n, liu2016n}.
Recent experimental realizations of monolayer and few-layer magnetic van der Waals crystals~\cite{gong2017n, huang2017n} have triggered a new wave of interest in the field of (quasi--) 2D magnets.
In particular, the interplay of dimensionality and magnetism, which can lead to useful magnetic properties,
is now being tested experimentally, calling for more efforts from the theory side.

Intrinsic ferromagnetic order has been shown to persist in mechanically exfoliated bilayers of Cr$_2$Ge$_2$Te$_6$~\cite{gong2017n} and monolayers of CrI$_3$~\cite{huang2017n}.
The bulk crystals of these materials consist of weakly van der Waals coupled layers, in which magnetic Cr atoms form a honeycomb lattice with edge-sharing octahedral coordination.
Inspired by these experimental discoveries, here we theoretically investigate the low-temperature properties of a ferromagnetic monolayer consisting of exchange-coupled spins forming a 2D honeycomb structure, as well as its few-layer (quasi--2D) and bulk (3D) versions, with a weak exchange coupling between the layers.
In this way, we explore the effect of dimensionality on the temperature dependence of magnetic ordering.

The standard argument demonstrating the extraordinary nature of magnetism in 2D is due to the Mermin-Wagner theorem excluding ferromagnetic order at any finite temperature in a 2D isotropic Heisenberg model.
However, even a tiny interaction breaking the rotational symmetry may stabilize ferromagnetic order in such a system at finite temperatures.
This argument testifies to the enhanced importance of anisotropic interactions in 2D.
In 3D, on the other hand, the Mermin-Wagner restriction is lifted due to the larger phase space, and the effect of anisotropic interactions is reduced.
Our prime interest here is the quantitative investigation of the cross-over from 2D to 3D realized by the few-layer magnetic systems mentioned above.

Here we adopt the ferromagnetic Heisenberg model of spins $S$ on the honeycomb lattice, including a single-ion anisotropy.
We treat this model by employing a self-consistently renormalized spin-wave theory (SRSWT)~\cite{bloch1962prl, loly1971jpcs, rastelli1974jpcs, pini1981jpcs, li2018jmmm}, which is an extension of the standard spin-wave analysis of the ferromagnet~\cite{dyson1956pr}.
The SRSWT is derived from the non-linear spin-wave theory, resulting from either applying the Dyson-Maleev transformation~\cite{dyson1956pr, maleev1958jetp} or using the Holstein-Primakoff transformation~\cite{holstein1940pr} with the subsequent truncation of the Hamiltonian up to quartic operator terms.
After either of these transformations, the quartic operator terms are treated in a Hartree-Fock-like decoupling approximation.
Within the Hartree-Fock decoupling, the Dyson-Maleev and the truncated Holstein-Primakoff transformation results coincide~\cite{liu1992jpcm}, yielding a quadratic Hamiltonian involving renormalization factors.
These temperature-dependent renormalization factors encapsulate the interaction-induced softening of the spin-wave spectrum.
The resulting quadratic Hamiltonian is then solved self-consistently.

In the majority of previous works on quasi-2D layered systems, a 3D Fourier transform is employed, including along the direction of layer stacking~\cite{kopietz1992prl, liu1992jpcm, irkhin1999prb, li2018jmmm}.
Even though the 3D Fourier transform facilitates the analytical treatment, it implies periodic boundary conditions in the direction of layer stacking, violating the difference between the surface and bulk layers.
However, for few-layer systems such treatment can be quite detrimental because these systems are intrinsically inhomogeneous and should be described with parameters that are different at the surface and the interior of the system.
In contrast, the SRSWT developed here is quite general and can be applied to systems with inequivalent sublattices.
This is utilized in the analysis of few-layered compounds where the layers are treated as separate sublattices~\cite{corciovei1963pr}.
Treating layers as separate sublattices allows us to account for the difference between the surface and bulk layers.
Moreover, it provides a natural way to resolve bulk magnon modes from the surface ones, which are known to occur in thin magnetic films~\cite{wolfram1972pss}.
The price we pay is the complication of dealing with a $2L\times 2L$ spin-wave Hamiltonian, where $L$ is the number of layers.
The $2L$ magnon eigenmodes of this Hamiltonian do not seem to be attainable analytically because of the lack of periodic boundary conditions along the layer-stacking direction.
Therefore, we solve the eigenvalue problem numerically at each point of the 2D momentum space.
In this way, results for systems involving up to seven intra-layer and inter-layer exchange couplings are found.
Although our approach is applicable to the general kind of inhomogeneity, we maintain a simple picture where the magnetic exchange and single-ion anisotropy parameters are the same throughout the system.
Even in this simplified picture, the surface layers are different from the bulk ones because of the difference in lattice coordination.

For Cr$_2$Ge$_2$Te$_6$ and CrI$_3$, different groups have calculated the exchange and magnetocrystalline anisotropy values from various ab initio methods~\cite{li2018jmmm, wang2016epl, besbes2019prb, torelli20182m, zhang2015jmcc, ke2021ncm}.
Notably, for CrI$_3$, calculation results of the same quantities are quite different.
Using the reported exchange and anisotropy values, we calculate the temperature-dependent magnetization $M(T)$ within our SRSWT and systematically compare the results.
We find that different parameter sets reported throughout the literature lead to the very different behavior of $M(T)$ and different Curie temperatures $\tc$, at which $M(T)$ vanishes.
This result indicates that more reliable methods for calculating effective magnetic interaction, and particularly for its anisotropic component, are desirable.

\section{Theory}

The two-dimensional honeycomb lattice can be viewed as a triangular lattice of unit cells, with two sites per unit cell.
Consequently, we label the sites of the honeycomb lattice by the pair, $(\mathbf{r} \nu)$, where $\mathbf{r}$ denotes the position of the unit cell and $\nu=1, 2$ labels the magnetic Cr sites within the unit cell.
We consider a lattice of atomic spins $S$, interacting with the Hamiltonian
\begin{equation}
\label{Hsys} H= \frac 12 \sum_{\mathbf{r} \nu} \sum_{\mathbf{r}'
\nu'}J_{\nu \nu'}^{\mathbf{r} \mathbf{r}'} \mathbf{S}_{\mathbf{r} \nu} \mathbf{S}_{\mathbf{r}' \nu'} -\sum_{\mathbf{r} \nu } \bigl [A (S^z_{\mathbf{r} \nu})^2 +g\mu_B B S^z_{\mathbf{r} \nu}\bigr ],
\end{equation}
where $\mathbf{S}_{\mathbf{r} \nu} = (S^x_{\mathbf{r} \nu}, S^y_{\mathbf{r} \nu}, S^z_{\mathbf{r} \nu})$ is the spin operator at the site $(\mathbf{r} \nu)$, $J_{\nu \nu'}^{\mathbf{r} \mathbf{r}'}$ is the Heisenberg exchange coupling between atomic spins at sites $(\mathbf{r} \nu)$ and $(\mathbf{r}' \nu')$, $A$ is the single-ion anisotropy along the $z$ direction (the direction normal to the plane of the atomic layer), $g$ is the Land\'{e} g-factor, $\mu_B$ is the Bohr magneton, and $B$ is the external magnetic field along the $z$ direction.
The ground state is the ferromagnetic state along the $z$ direction, implying positive $A$ (easy-axis anisotropy) and predominantly negative $J_{\nu \nu'}^{\mathbf{r} \mathbf{r}'}$.

A spin wave theory for the above model can be derived by using either the Dyson-Maleev transformation~\cite{dyson1956pr, maleev1958jetp}, or the Holstein-Primakoff transformation~\cite{holstein1940pr} followed by the truncation of terms higher than quartic.
In the approximation that follows, the two transformations yield equivalent results.
These transformations map spin operators $\mathbf{S}_{\mathbf{r} \nu}$ onto bosonic creation-annihilation operators $a^\dagger_{\mathbf{r} \nu}$, $a_{\mathbf{r} \nu}$, with commutation relations, $[a_{\mathbf{r} \nu}, a^\dagger_{\mathbf{r}' \nu'}] =\delta_{\mathbf{r} \mathbf{r}'}\delta_{\nu \nu'}$, $[a_{\mathbf{r} \nu}, a_{\mathbf{r}' \nu}'] =[a^\dagger_{\mathbf{r} \nu}, a^\dagger_{\mathbf{r}' \nu'}]=0$, as
\begin{eqnarray}
  \label{HPtrans}
&&S^z_{\mathbf{r} \nu}=S-a^\dagger_{\mathbf{r} \nu}a_{\mathbf{r} \nu},\nonumber \\
&&S^+_{\mathbf{r} \nu}=\sqrt{2S}\left (1 -\frac{
a^\dagger_{\mathbf{r} \nu} a_{\mathbf{r} \nu}}{2S} \right )^\xi
a_{\mathbf{r} \nu},\nonumber \\
&&S^-_{\mathbf{r} \nu}= \sqrt{2S}\, a^\dagger_{\mathbf{r} \nu}
\left (1 -\frac{ a^\dagger_{\mathbf{r} \nu} a_{\mathbf{r}
\nu}}{2S} \right )^{1- \xi},
\end{eqnarray}
where $\xi=1$ for the Dyson-Maleev and $\xi=1/2$ for the Holstein-Primakoff transformations.
For a system of $N$ unit cells under periodic boundary condition, it is convenient to introduce the Fourier transforms,
\begin{equation}
  \label{FT}
  a^\dagger_{\mathbf{r} \nu}
  =\frac1{\sqrt{N}}\sum_{\mathbf{k}}e^{-i\mathbf{k}\cdot \mathbf{r}}
  b^\dagger_{\mathbf{k} \nu},\quad a_{\mathbf{r} \nu}
  =\frac1{\sqrt{N}}\sum_{\mathbf{k}}e^{i\mathbf{k}\cdot \mathbf{r}}
  b_{\mathbf{k} \nu},
\end{equation}
where $\mathbf{k}$ runs over the first Brillouin zone of the triangular lattice of unit cells, and operators $b^\dagger_{\mathbf{k} \nu}$, $b_{\mathbf{k} \nu}$, satisfy the bosonic commutation relations, $[b_{\mathbf{k} \nu}, b^\dagger_{\mathbf{k}' \nu'}] =\delta_{\mathbf{k} \mathbf{k}'}\delta_{\nu \nu'}$, $[b_{\mathbf{k} \nu}, b_{\mathbf{k}' \nu}'] =[b^\dagger_{\mathbf{k} \nu}, b^\dagger_{\mathbf{k}' \nu'}]=0$.
The Bravais lattice structure of unit cells ensures the relation, $\sum_{\mathbf{r}}e^{i(\mathbf{k} -\mathbf{k}')\cdot \mathbf{r}}= N\delta_{\mathbf{k} \mathbf{k}'}$.

After applying the mapping \req{HPtrans} and expanding the result with respect to large $S$, the ferromagnetic ground state energy emerges as the term independent of the Bose operators,
\begin{equation}
\label{E0}
  E_0=2N\left( \tilde{J} S^2 -AS^2 -g\mu_B BS \right),
\end{equation}
where $\tilde{J}= \sum_{\mathbf{r}' \nu'} J^{\mathbf{r} \mathbf{r}'}_{\nu \nu'}$.
The non-interacting magnon Hamiltonian, $H_2$, is further found as the part quadratic in the Bose operators.
In terms of the Fourier representation of exchange couplings
\begin{equation}
\label{FTintro}
  J^{\mathbf{k}}_{\nu \nu'}=\sum_{\mathbf{r}'}
  e^{-i\mathbf{k} \cdot \mathbf{R}_{\nu \nu'}^{\mathbf{r} \mathbf{r}'}}
  J^{\mathbf{r} \mathbf{r}'}_{\nu \nu'},
\end{equation}
where $\mathbf{R}_{\nu \nu'}^{\mathbf{r} \mathbf{r}'}$ is the vector connecting sites $(\mathbf{r} \nu)$ and $(\mathbf{r}' \nu')$ [e.g., $\mathbf{R}_{\nu \nu}^{\mathbf{r} \mathbf{r}'} =\mathbf{r}-\mathbf{r}'$], the explicit form of $H_2$ is
\begin{eqnarray}
\label{H0expl}
  H_2 = &&\frac S2 \sum_{\mathbf{k} \nu \nu'}\bigl(
  J^{\mathbf{k}}_{\nu \nu'} b^\dagger_{\mathbf{k} \nu} b_{\mathbf{k} \nu'}
  +J^{-\mathbf{k}}_{\nu \nu'} b^\dagger_{\mathbf{k} \nu'}
  b_{\mathbf{k} \nu}\bigr) \nonumber \\
  && -\sum_{\mathbf{k} \nu}\left (S\tilde{J} -2SA
  -g\mu_BB \right )b^\dagger_{\mathbf{k} \nu} b_{\mathbf{k} \nu}.
\end{eqnarray}
Note that \req{H0expl}, as well as the entire treatment that follows, exploits the independence of $J^{\mathbf{k}}_{\nu \nu'}$ defined by \req{FTintro} on $\mathbf{r}$.
This property is due to the absence of boundaries, resulting from the periodic boundary condition.
Otherwise, if the system has boundaries, the sum in \req{FTintro} depends on whether $\mathbf{r}$ is an internal unit cell or it is located at a boundary, where some of its neighboring sites are missing.

In the absence of dipolar interaction, the next terms in the large - $S$ expansion are the four-boson terms comprising the spin-wave interaction Hamiltonian, $H_4$.
The form of $H_4$ depends on what specific spin-boson mapping is applied, one obvious difference being that the Holstein-Primakoff result is Hermitian, unlike the Dyson-Maleev one.
The Holstein-Primakoff mapping yields
\begin{widetext}
\begin{eqnarray}
\label{H1}
  H_4 = &&\frac 1{4N}\! \sum_{\mathbf{k}_i, \nu} \Biggl \{\!
  \sum_{\nu'}\!
  \left[ 2J^{\mathbf{k}_1 -\mathbf{k}_3}_{\nu \nu'} b^\dagger_{\mathbf{k}_1 \nu}
  b^\dagger_{\mathbf{k}_2 \nu'} b_{\mathbf{k}_3 \nu}
  b_{\mathbf{k}_4 \nu'} - J^{\mathbf{k}_1}_{\nu \nu'} b^\dagger_{\mathbf{k}_1 \nu}
  b^\dagger_{\mathbf{k}_2 \nu'} b_{\mathbf{k}_3 \nu'}
  b_{\mathbf{k}_4 \nu'}  - J^{\mathbf{k}_4}_{\nu \nu'} b^\dagger_{\mathbf{k}_1 \nu}
  b^\dagger_{\mathbf{k}_2 \nu} b_{\mathbf{k}_3 \nu} b_{\mathbf{k}_4
  \nu'} \right ]  \nonumber \\
  && -4 A b^\dagger_{\mathbf{k}_1 \nu}
  b^\dagger_{\mathbf{k}_2 \nu} b_{\mathbf{k}_3 \nu}
  b_{\mathbf{k}_4 \nu} \Biggr \}\delta_{
  \mathbf{k}_1+\mathbf{k}_2,\mathbf{k}_3 +\mathbf{k}_4}
  -\sum_{\mathbf{k}\nu } A b^\dagger_{\mathbf{k} \nu}
  b_{\mathbf{k} \nu},
\end{eqnarray}
\end{widetext}
where the last, linear in $b^\dagger_{\mathbf{k} \nu} b_{\mathbf{k} \nu}$ term, associated with the single-ion anisotropy $A$, originates from the commutation relations between the bosonic operators.
Note that the Dyson-Maleev mapping leads to \req{H1} with the second, $\propto J^{\mathbf{k}_1}_{\nu \nu'}$ term doubled and the third, $\propto J^{\mathbf{k}_4}_{\nu \nu'}$ term missing.

The essential approximation that leads to the renormalized spin-wave theory is the Hartree-Fock-like decoupling of four-boson terms,
\begin{eqnarray}
\label{HF4bt}
  &&b^\dagger_{\mathbf{k}_1 \nu_1} b^\dagger_{\mathbf{k}_2 \nu_2}
  b_{\mathbf{k}_3 \nu_3} b_{\mathbf{k}_4 \nu_4}\approx  \\
  && \langle b^\dagger_{\mathbf{k}_1 \nu_1} b_{\mathbf{k}_3 \nu_3}\rangle
  b^\dagger_{\mathbf{k}_2 \nu_2} b_{\mathbf{k}_4 \nu_4}
  + \langle b^\dagger_{\mathbf{k}_1 \nu_1} b_{\mathbf{k}_4 \nu_4}\rangle
  b^\dagger_{\mathbf{k}_2 \nu_2} b_{\mathbf{k}_3 \nu_3} \nonumber \\
  &&+  \langle b^\dagger_{\mathbf{k}_2 \nu_2} b_{\mathbf{k}_3 \nu_3}\rangle
  b^\dagger_{\mathbf{k}_1 \nu_1} b_{\mathbf{k}_4 \nu_4}
  +  \langle b^\dagger_{\mathbf{k}_2 \nu_2} b_{\mathbf{k}_4 \nu_4}\rangle
  b^\dagger_{\mathbf{k}_1 \nu_1} b_{\mathbf{k}_3 \nu_3}.\quad \nonumber
\end{eqnarray}
Here we skip scalar terms which do not affect the spin-wave dynamics.
Furthermore, in \req{HF4bt} we keep only terms containing averages with coinciding $\mathbf{k}$-indices (the so-called diagonal terms~\cite{bloch1962prl, li2018jmmm}), i.e., we utilize
\begin{equation}
\label{bav}
  \langle b^\dagger_{\mathbf{k} \nu} b_{\mathbf{k}' \nu'}\rangle =
  \delta_{\mathbf{k} \mathbf{k}'} \langle b^\dagger_{\mathbf{k} \nu} b_{\mathbf{k}
  \nu'}\rangle.
\end{equation}
This relation can be justified by noting that the resulting magnon modes are diagonal in $\mathbf{k}$, involving no $\mathbf{k}$ -- $\mathbf{k}'$ mixing.
Note, however, that both $H_2$ and $H_4$ are essentially non-diagonal in $\nu$ - indices, and the resulting magnon modes are coherent superpositions of $b_{\mathbf{k} \nu}$ - bosons with different $\nu$ - indices.
Therefore, the two-boson average in \req{bav} is essentially non-diagonal in the $\nu$ - indices.
In Ref.~[\onlinecite{li2018jmmm}] this fact is ignored, and two-boson averages non-diagonal in $\nu$ - indices are eliminated.

By applying the above mean-field approximation to \req{H1} or its Dyson-Maleev counterpart and combining the result with \req{H0expl}, for the interacting spin-wave Hamiltonian $H_2+H_4$ we get the mean-field expression (the renormalized spin-wave Hamiltonian),
\begin{equation}
\label{HMF}
  H_R= \sum_{\mathbf{k} \nu \nu'}\mathcal{J}_{\nu \nu'}({\mathbf{k}})
  b^\dagger_{\mathbf{k} \nu} b_{\mathbf{k} \nu'}
  + \sum_{\mathbf{k} \nu} \mathcal{L}_\nu({\mathbf{k}})
  b^\dagger_{\mathbf{k} \nu} b_{\mathbf{k} \nu},
\end{equation}
with the coefficients given by
\begin{widetext}
\begin{eqnarray}
\label{coeffs}
  \mathcal{J}_{\nu \nu'}({\mathbf{k}}) = &&
  \left (S - \frac 1{2N} \sum_{\mathbf{k}'} \left[ n_{\mathbf{k}' \nu \nu} +n_{\mathbf{k}' \nu' \nu'}\right] \right)
  J^{\mathbf{k}}_{\nu \nu'} + \frac 1N \sum_{\mathbf{k}'} J^{\mathbf{k}-\mathbf{k}'}_{\nu \nu'}n_{\mathbf{k}' \nu' \nu},   \nonumber \\
  \mathcal{L}_\nu({\mathbf{k}}) = && g\mu_B B -\left (1-2S + \frac 4N \sum_{\mathbf{k}'} n_{\mathbf{k}' \nu \nu} \right)A
  - \sum_{\nu'}\left (S - \frac 1N \sum_{\mathbf{k}'}
  n_{\mathbf{k}' \nu' \nu'} \right)J^{\mathbf{k}=0}_{\nu \nu'} \nonumber \\
  &&-\frac 1{2N} \sum_{\mathbf{k}' \nu'} \left ( J^{-\mathbf{k}'}_{\nu \nu'} n_{\mathbf{k}' \nu' \nu}
  +J^{-\mathbf{k}'}_{\nu' \nu} n_{\mathbf{k}' \nu \nu'} \right),
\end{eqnarray}
\end{widetext}
where two-boson thermal averages, $n_{\mathbf{k} \nu \nu'} = \langle b^\dagger_{\mathbf{k} \nu} b_{\mathbf{k} \nu'} \rangle$, are introduced.
Through \req{coeffs}, these averages define the temperature-dependent renormalization factors encapsulating spin-wave interaction effects at the Hartree-Fock level.

At this point, we note that the Hamiltonian \req{HMF} is quite universal in that it is suitable to any system of spins $S$ on a generic Bravais lattice of $N$ unit cells under the periodic boundary condition, with an arbitrary number of magnetic sites, $n$, in the unit cell.
To further retain the universal form, we introduce the structure factors as
\begin{equation}
\label{univgamma}
  \gamma^\rho_{\nu \nu'}(\mathbf{k})= J_\rho\sum_ {\mathbf{u}^\rho_{\nu \nu'}}
  e^{i\mathbf{k}\cdot\mathbf{u}^\rho_{\nu \nu'}},
\end{equation}
where $\rho$ enumerates the non-zero exchange couplings and $\mathbf{u}^\rho_{\nu \nu'}$ run over the links between a given spin on the sublattice $\nu$ and those spins on the sublattice $\nu'$ that are coupled to the given one by the exchange $J_\rho$.
Then, for Fourier transforms of exchange coupling \req{FTintro} we get
\begin{equation}
\label{univJk}
  J^{\mathbf{k}}_{\nu \nu'}= \sum_{\rho \in [\nu\nu']}
  \gamma^\rho_{\nu \nu'}(\mathbf{k}), \qquad \nu, \nu' =1,\cdots,n,
\end{equation}
with $\rho \in [\nu\nu']$ meaning that the spins on sublattices $\nu$ and $\nu'$ are coupled by the exchange $J_\rho$.
Furthermore, by noting that $n_{\mathbf{k} \nu \nu}$ is the average number of bosonic excitations on the sublattice $\nu$, we introduce the sublattice spin polarization,
\begin{equation}
\label{slavmag}
  \bar{S}_\nu = S -\frac 1N\sum_{\mathbf{k}} n_{\mathbf{k} \nu
  \nu},\qquad \nu =1,\cdots,n.
\end{equation}
Additional thermodynamic quantities are introduced by the relation,
\begin{equation}
\label{univfis}
  f^\rho_{\nu \nu'} = \frac 1N\sum_{\mathbf{k}}
  \frac{\gamma^\rho_{\nu \nu'}(\mathbf{-k})} {\gamma^\rho_{\nu \nu'}(0)}
  n_{\mathbf{k} \nu' \nu }, \quad \rho \in [\nu \nu'].
\end{equation}
The physical meaning of $f^\rho_{\nu \nu'}$ becomes apparent from the real-space expression,
\begin{equation}
\label{fjrs}
  f^\rho_{\nu \nu'}= \frac 1{z^\rho_{\nu \nu'}}\sum_ {\mathbf{u}^\rho_{\nu \nu'}}
  \langle a^\dagger_{\mathbf{r} \nu} a_{\mathbf{r} \nu + \mathbf{u}^\rho_{\nu \nu'}}
  \rangle, \quad \rho \in [\nu \nu'],
\end{equation}
where $z^\rho_{\nu \nu'} =\sum_ {\mathbf{u}^\rho_{\nu \nu'}} 1$ is the number of $\mathbf{u}^\rho_{\nu \nu'}$ (the so-called coordination number).
Thus, $f^\rho_{\nu \nu'}$ is the short-range bosonic correlation between the sites on sublattices $\nu$ and $\nu'$ exchange-coupled through $J_\rho$.
With these notations, we rewrite \req{coeffs} as
\begin{eqnarray}
\label{univcoeffs}
  \mathcal{J}_{\nu \nu'}({\mathbf{k}}) &&=
  \sum_{\rho\in[\nu \nu']}\left \{\frac 12  \left( \bar{S}_\nu +\bar{S}_{\nu'} \right) +f^\rho_{\nu \nu'}
  \right\}\gamma^\rho_{\nu \nu'}(\mathbf{k}),   \nonumber \\
  \mathcal{L}_\nu({\mathbf{k}}) &&= g\mu_B B -\left (2S +1 - 4\bar{S}_\nu
  \right)A \nonumber \\
  && - \sum_{\nu'}\sum_{\rho\in[\nu \nu']}
  \left \{\bar{S}_{\nu'} + \text{Re}(f^\rho_{\nu \nu'}) \right\}\gamma^\rho_{\nu
  \nu'}(0).
\end{eqnarray}

The self-consistency equations are found by expressing the thermodynamic quantities $\bar{S}_\nu$ and $f^\rho_{\nu \nu'}$ through the Hamiltonian defined in \req{HMF}.
To this end, we consider the creation-annihilation operators $\alpha^\dag_{\mathbf{k}, \sigma}$, $\alpha_{\mathbf{k}, \sigma}$ of magnon eigenmodes of the Hamiltonian $H_R$, where $\sigma$ labels the $n$ magnon branches, and note that $\alpha^\dag_{\mathbf{k}, \sigma}$ and $\alpha_{\mathbf{k}, \sigma}$  are linear combinations of $b^\dag_{\mathbf{k}, \nu}$ and $b_{\mathbf{k}, \nu}$, respectively.
One has
\begin{equation}
\label{bviamm}
  b_{\mathbf{k}\nu} = \sum_{\sigma=1}^n  \left[\Lambda_\mathbf{k}\right]_{\nu \sigma}
  \alpha_{\mathbf{k}\sigma},
\end{equation}
and the corresponding complex conjugate relation between $b^\dagger_{\mathbf{k} \nu}$ and $\alpha^\dagger_{\mathbf{k} \sigma}$, where $\Lambda_\mathbf{k}$ is the eigenvector matrix that diagonalizes $H_R$, and brackets meaning matrix elements.
From \req{bviamm}, its Hermitian conjugate, and the Bose-Einstein relation $\langle \alpha^\dagger_{\mathbf{k}\sigma} \alpha_{\mathbf{k}\sigma} \rangle=1/(e^{\beta E_{\sigma} (\mathbf{k})} -1)$, where $\beta = 1/k_\mathrm{B}T$ is the inverse temperature and $E_{\sigma}(\mathbf{k})$ is the magnon dispersion of $H$, one finds
\begin{equation}
\label{n21}
 n_{\mathbf{k} \nu \nu'}\equiv \langle b^\dagger_{\mathbf{k}\nu} b_{\mathbf{k}\nu'} \rangle
 = \sum_{\sigma}\frac {\left[\Lambda_\mathbf{k}^*\right]_{\nu \sigma}
 \left[\Lambda_\mathbf{k}\right]_{\nu' \sigma}}{e^{\beta
  E_{\sigma}(\mathbf{k})}-1}.
\end{equation}
Thus, we arrive at the self-consistency equations,
\begin{eqnarray}
\label{scSs}
  &&\bar{S}_\nu = S -\frac 1N\sum_{\mathbf{k},\sigma}
  \frac {\left[\Lambda_\mathbf{k}^*\right]_{\nu \sigma}
  \left[\Lambda_\mathbf{k}\right]_{\nu \sigma}}{e^{\beta
  E_{\sigma}(\mathbf{k})}-1}, \\
\label{scfs}
  &&f^\rho_{\nu \nu'} = \frac 1N
  \sum_{\mathbf{k}, \sigma} \frac{\gamma^\rho_{\nu \nu'}(\mathbf{-k})}
  {\gamma^\rho_{\nu \nu'}(0)}
  \frac {\left[\Lambda_\mathbf{k}^*\right]_{\nu' \sigma}
  \left[\Lambda_\mathbf{k}\right]_{\nu \sigma}}{e^{\beta
  E_{\sigma}(\mathbf{k})}-1},
\end{eqnarray}
which are to be solved numerically, for the average magnetization $\bar{S} =\sum_\nu \bar{S}_\nu/n$.
For systems considered below we find that the self-consistency equations have solutions with real $f^\rho_{\nu \nu'}$, entailing the symmetry, $\mathcal{J}^*_{\nu \nu'}({\mathbf{k}})= \mathcal{J}_{\nu \nu'}(-{\mathbf{k}})$.
While this is natural for monolayer and bulk systems with equivalent sublattices, for layered systems with inequivalent surface and bulk sublattices it is less intuitive.

Note in passing that alternatively to employing the Hartree-Fock decoupling \req{HF4bt}, the above SRSWT could be derived from the Feynman-Peierls-Bogoliubov variational principle~\cite{huber1969chapter}, in exactly the same form.
In addition, the SRSWT is equivalent to the summation of all bubble diagrams for the self-energy~\cite{loly1971jpcs}.

\begin{figure}[t]
\centerline{\includegraphics[width=\linewidth,angle=0,clip]{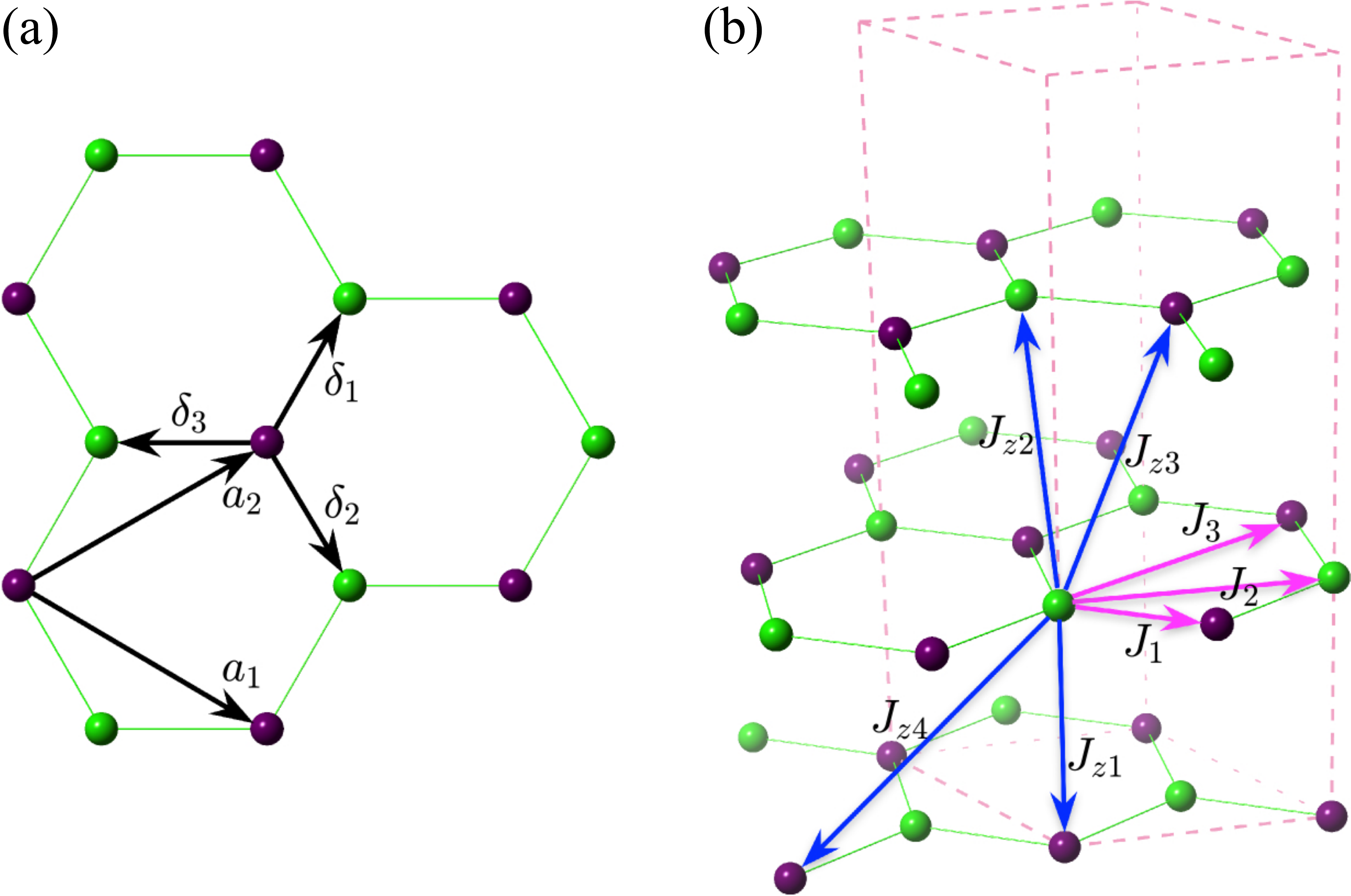}}
\caption{(Color online) Schematic crystal structure of a layered system with magnetic atoms forming honeycomb structure in each 2D layer, and ABC-type layer stacking.
The two in-plane sublattices sites are depicted with green and purple balls.
(a) The 2D honeycomb structure is characterized by the unit cell primitive vectors, $\mathbf{a}_1$ and $\mathbf{a}_2$.
The vectors $\boldsymbol{\delta}_1$, $\boldsymbol{\delta}_2$, and $\boldsymbol{\delta}_3$ link the first nearest neighbor sites.
(b) Magnetic interaction in each layer is characterized by the three non-zero exchange couplings between the first, second, and third nearest neighbor spins, denoted by $J_1$, $J_2$, and $J_3$, respectively (magenta arrows).
The vdW-bonded layers are displaced upon stacking so that the magnetic ion in one layer is directly over the center of a hexagon of one of the two adjacent layers, forming a rhombohedral ABC stacking sequence.
The inter-layer exchange couplings, $J_{z1}$, $J_{z2}$, $J_{z3}$, and $J_{z4}$, are illustrated with blue arrows.}
\label{fewlayers}
\end{figure}

In what follows, we apply the above approach to layered systems with magnetic atoms forming honeycomb structure in each 2D layer, and ABC-type layer stacking.
We begin our consideration with the monolayer of intra-layer exchange couplings $J_1$, $J_2$, and $J_3$ between the first, second, and third nearest neighbor spins, respectively [see \rfig{fewlayers}(a)].
Subsequently, we address the bulk (3D) and the few-layer (quasi-2D) cases, with up to four additional non-zero inter-layer exchange couplings, $J_{z1}$, $J_{z2}$, $J_{z3}$, and $J_{z4}$, specified in \rfig{fewlayers}(b).

\subsection{Monolayer (2D) system}

For the honeycomb monolayer with two sublattices and three non-zero exchange couplings $J_1$, $J_2$, and $J_3$, shown in \rfig{fewlayers}(a), one has the index $\rho$ running over the three values $1$, $2$, $3$, and sublattice indices taking two values; $\nu$, $\nu'=1,2$.
Thus, there are three structure factors,
\begin{equation}
\label{2Dgammas}
  \gamma_1(\mathbf{k})\equiv \gamma^1_{12}(\mathbf{k}),\quad
  \gamma_2(\mathbf{k})\equiv \gamma^2_{\nu \nu}(\mathbf{k}),\quad
  \gamma_3(\mathbf{k})\equiv \gamma^3_{12}(\mathbf{k}),
\end{equation}
and, due to the equivalence of the sublattices, only four independent thermodynamic quantities:
\begin{eqnarray}
\label{2DSf}
 &&\bar{S}=\bar{S}_1=\bar{S}_2,\nonumber \\
 &&f_1\equiv f^1_{1 2}, \quad f_2\equiv f^2_{\nu \nu},\quad
 f_3\equiv f^3_{1 2}.
\end{eqnarray}
Using Eqs.~(\ref{HMF}), (\ref{univcoeffs}), (\ref{2Dgammas}), and
(\ref{2DSf}), we find the Hamiltonian of the monolayer,
\begin{equation}
\label{H2D} H_{2D}= \sum_{\mathbf{k}}\bigl (
b^\dagger_{\mathbf{k}1}\, b^\dagger_{\mathbf{k}2} \bigr) \left (
\begin{array}{cc}
\mathcal{A}(\mathbf{k}) &\mathcal{B}(\mathbf{k}) \\
\mathcal{B}^*(\mathbf{k}) &\mathcal{A}(\mathbf{k})
\end{array} \right ) \left ( \begin{array}{c}
b_{\mathbf{k}1} \\
b_{\mathbf{k}2}
\end{array}
\right ),
\end{equation}
where
\begin{eqnarray}
\label{coeffsAB}
  \mathcal{A}({\mathbf{k}}) = && g\mu_B B -\left (2S+1 -4\bar{S} \right)A  \nonumber \\
   &&+\left( \bar{S} +f_2 \right)\gamma_2(\mathbf{k})
   - \sum_{\rho=1}^3\!\! \left( \bar{S} + f_\rho \right)\gamma_\rho(0),   \nonumber \\
  \mathcal{B}({\mathbf{k}}) = && \left (\bar{S} +f_1 \right )\gamma_1(\mathbf{k})
  +\left (\bar{S} +f_3 \right )\gamma_3(\mathbf{k}).
\end{eqnarray}
The dispersion relation following from \req{H2D} reads
\begin{equation}
\label{Epm}
  E_{\pm}(\mathbf{k})=  \mathcal{A}({\mathbf{k}})
  \pm|\mathcal{B}({\mathbf{k}})|,
\end{equation}
with the in-phase acoustic (labeled by $-$) and out-of-phase optical (labeled by $+$) branches.
It is also straightforward to find the explicit form of $\Lambda_\mathbf{k}$ diagonalizing the matrix \req{H2D},
\begin{equation}
\label{Lambexp}
  \Lambda_\mathbf{k} = \frac 1{\sqrt{2}}\left ( \begin{array}{cc}
  1 &-e^{i\phi_\mathbf{k}} \\
  e^{-i\phi_\mathbf{k}} &1
  \end{array} \right ),
\end{equation}
with the phase, $\phi_\mathbf{k}= \arg \left [\mathcal{B}({\mathbf{k}}) \right ]$.
Putting together Eqs.~(\ref{scSs}), (\ref{scfs}) and (\ref{Lambexp}),
we arrive at the self-consistency equations
\begin{eqnarray}
\label{avmag}
  &&\bar{S}=  S-\frac 1{2N} \sum_{\sigma=\pm}
  \sum_{\mathbf{k}}\frac 1{e^{\beta E_{\sigma}(\mathbf{k})}-1}, \\
\label{f1f3expl}
  &&f_\rho = \frac 1{2N \gamma_\rho(0)}\sum_{\sigma=\pm}\sum_{\mathbf{k}}
  \frac {\sigma\gamma_\rho(\mathbf{k}) e^{-i\phi_\mathbf{k}}}{e^{\beta
  E_{\sigma}(\mathbf{k})}-1},
  \quad \rho=1, 3, \quad \\
\label{f2expl}
  &&f_2 =\frac 1{2N \gamma_2(0)}\sum_{\sigma=\pm}\sum_{\mathbf{k}}
  \frac {\gamma_2(\mathbf{k})}{e^{\beta
  E_{\sigma}(\mathbf{k})}-1}.
\end{eqnarray}

Equations (\ref{avmag}) -- (\ref{f2expl}) for $\bar{S}$ and $f_\rho$, together with \req{Epm} for $E_{\pm}(\mathbf{k})$, constitute a closed set of equations which is solved for the average magnetization, $\bar{S}$.
This is done numerically, by utilizing the following iterative algorithm.
At the initial step, the four input values $\bar{S}^{(0)}= S$ and $f^{(0)}_\rho= 0$, are plugged into Eqs.~(\ref{coeffsAB}) and (\ref{Epm}) to find the initial spectrum $E^{(0)}_{\pm}(\mathbf{k})$ and phase $\phi^{(0)}_\mathbf{k}$.
Then, $E^{(0)}_{\pm}(\mathbf{k})$ and $\phi^{(0)}_\mathbf{k}$ are used in the right-hand sides of Eqs.
(\ref{avmag}) -- (\ref{f2expl}), to calculate the next-order four values $\bar{S}^{(1)}$ and $f^{(1)}_\rho$.
Likewise, at the $i$-th iteration step, four input values $\bar{S}^{(i-1)}$ and $f^{(i-1)}_\rho$ are fed to Eqs.
(\ref{coeffsAB}), (\ref{Epm}), yielding the next-order spectrum and phase, $E^{(i-1)}_{\pm}(\mathbf{k})$ and $\phi^{(i-1)}_\mathbf{k}$, which are subsequently used in Eqs.
(\ref{avmag}), (\ref{f2expl}), and (\ref{f1f3expl}) to find the four output values, $\bar{S}^{(i)}$ and $f^{(i)}_\rho$.
This procedure is repeated until the four input and output values converge within a desired accuracy, producing the value of $\bar{S}$.

\subsection{Bulk (3D) system}
The SRSWT is easily generalized to the bulk system of ABC-type layer stacking along the $z$ direction, provided that periodic boundary condition is imposed in the $z$ direction, likewise the in-plane directions.
This generalization is facilitated by the fact that introducing the third component of the wavevector, corresponding to the new spatial direction, retains the two-sublattice structure independently of the number of layers.
Thus, for the bulk system with three intra-layer exchange couplings $J_1$, $J_2$, $J_3$, and four inter-layer exchange couplings $J_{z1}$, $J_{z2}$, $J_{z3}$, $J_{z4}$, shown in \rfig{fewlayers}, we have
eight inequivalent thermodynamic quantities; $\bar{S}$ and $f_\rho$, $\rho=1$, $2$, $3$, $z1$, $z2$, $z3$, $z4$.
The complete analysis of this case is presented in Appendix~\ref{App3D}.
Formally, this analysis follows the same steps as that of the monolayer.
A magnon dispersion relation, consisting of two branches, is analytically found as a function of thermodynamic quantities $\bar{S}$ and $f_\rho$.
In turn, these quantities are expressed in terms of the magnon dispersion, much like in Eqs.~(\ref{avmag}), (\ref{f1f3expl}), and (\ref{f2expl}).
This sets up a system of self-consistency equations which we solve numerically for $\bar{S}$.

\subsection{Few-layer (quasi-2D) system}

The principal difference of the few-layer system with ABC-type layer stacking along the $z$ direction from the bulk case discussed above is that the few-layer system has two surface layers, which are not equivalent to the inner layers, simply because magnetic atoms in surface layers have some missing neighbors.
Therefore, surface layers of the few-layer system violate periodic boundary conditions in the stacking direction, making the formal extension of the Fourier transform to the third direction inapplicable.
As a matter of fact, the physical difference of surface and bulk layers may even result in distinct values of exchange and single-ion anisotropy parameters.
The approach that follows is suitable for systems with different exchange and single-ion anisotropy parameters at different layers.
However, for the sake of simplicity, in our subsequent simulations we assume that interaction parameters are the same throughout the system.

In line with the long-known approach~\cite{corciovei1963pr}, we consider the few-layer system with $L$ layers comprised of $2L$ sublattices, two per layer.
The corresponding Hamiltonian, $H_L$, is given by \req{HMF}, with $\nu$ and $\nu'$ running over the values $1,\cdots, 2L$.
Utilizing the operator-valued vector,
\begin{equation}
\label{psi}
  \boldsymbol{\psi}^\dagger(\mathbf{k})=
  \left(b^\dagger_{\mathbf{k} 1}, b^\dagger_{\mathbf{k} 2},\cdots,
  b^\dagger_{\mathbf{k}\, (2L-1)}, b^\dagger_{\mathbf{k}\, (2L)}\right),
\end{equation}
where $\mathbf{k}$ is the two-dimensional wavevector, we write the Hamiltonian as
\begin{equation}
\label{HL}
  H_L=\boldsymbol{\psi}^\dagger(\mathbf{k})\hat{H}_\mathbf{k}
  \boldsymbol{\psi}(\mathbf{k}).
\end{equation}
To visualize the underlying layered structure, we represent $\hat{H}_\mathbf{k}$ in the form of an $L\times L$ matrix of $2\times 2$ entries,
\begin{equation}
\label{Hkmrx}
  \hat{H}_\mathbf{k}=\left (
  \begin{array}{ccccccc}
  \hat{h}_1 &\hat{b}_1 &0 &\cdots &0 &0 &0\\
  \hat{b}_1^\dag &\hat{h}_2 &\hat{b}_2 &\cdots &0 &0 &0\\
  0 &\hat{b}_2^\dag &\hat{h}_3 &\cdots &0 &0 &0\\
  \vdots &\vdots &\vdots &\ddots &\vdots &\vdots &\vdots\\
  0 &0 &0 &\cdots &\hat{h}_{L-2} &\hat{b}_{L-2} &0\\
  0 &0 &0 &\cdots &\hat{b}_{L-2}^\dag &\hat{h}_{L-1} &\hat{b}_{L-1}\\
  0 &0 &0 &\cdots &0 &\hat{b}_{L-1}^\dag &\hat{h}_L
  \end{array} \right ),
\end{equation}
with $\hat{h}_l$ corresponding to the $l$-th layer, where the layers are enumerated from bottom to top along the stacking direction.
As a consequence of the physical difference of surface and bulk layers,
$\hat{h}_l$ with $l=2, \cdots, L-1$ have identical structure,
different from that of $\hat{h}_1$ and $\hat{h}_L$ corresponding to the surface layers.
At the same time, all $\hat{b}$ - operators are structurally identical, reflecting the fact that, in the approximation we work, inter-layer interactions between all successive layers are uniform.
Explicit forms of $\hat{h}$ - and $\hat{b}$ - operators are given in Appendix~\ref{AppQ2D}.

Altogether, $\hat{H}_\mathbf{k}$ is a function of $2L$ sublattice spin polarizations $\bar{S}_\nu$, $4L$ intra-layer short-range correlations $f^\rho_{\nu \nu'}$ ($\rho =1$, $2$, $3$), and $(5L-5)$ inter-layer short-range correlations $f^\rho_{\nu \nu'}$ ($\rho =z1$, $z2$, $z3$, $z4$).
However, sublattices labeld by the indices $\nu$ and $2L+1-\nu$ are equivalent, and it is reasonable to expect that sublattice magnetizations on equivalent sublattices as well as short-range correlations between the equivalent pairs of sublattices are the same.
This reduces the total number of independent variables to $C_L=(11L-3)/2$ or $C_L=(11L-2)/2$ for odd or even $L$, respectively.
Accordingly, the self-consistency is a system of $C_L$ equations, which we solve numerically.

Because of the number of sublattices larger than two, analytical steps that followed Eqs.~(\ref{scSs}), (\ref{scfs}) in the two previous cases of monolayer and bulk systems are inaccessible for the quasi-2D system.
In particular, for generic $L$, analytical expressions are not available for magnon dispersion $E_\sigma(\mathbf{k})$ and transformation matrix $\Lambda_\mathbf{k}$, which is the matrix of eigenvectors of $\hat{H}_\mathbf{k}$.
Therefore, we extend the previous simulation procedure and include an extra step for numerical diagonalization of $\hat{H}_\mathbf{k}$, at each $\mathbf{k}$-point.
In other words, we address the self-consistency equations by solving the eigenvalue problem
\begin{equation}
\label{numdiag}
  \hat{H}_\mathbf{k}\Lambda_\mathbf{k} = \Lambda_\mathbf{k}\, \text{diag} \bigl[
  E_1(\mathbf{k}), \cdots , E_{2L}(\mathbf{k})\bigr],
\end{equation}
numerically.
Thus, at the $i$-th numerical iteration step, $C_L$ input parameters $\bar{S}^{(i-1)}_\nu$ and $f^{\rho (i-1)}_{\nu \nu'}$ are taken as arguments of $\hat{H}_\mathbf{k}$ to calculate the eigenmodes, $E^{(i-1)}_{\nu}(\mathbf{k})$, $\Lambda_\mathbf{k}^{(i-1)}$, which are subsequently used in Eqs.~(\ref{scSs}) and (\ref{scfs}) to find the output parameters, $\bar{S}^{(i)}_\nu$ and $f^{\rho (i)}_{\nu \nu'}$, until the results converge.
Because of the extra numerical diagonalization step, the simulation procedure for the quasi-2D system is much more demanding than those for the two previous cases of monolayer and bulk systems.

  \begin{table*}[ht]
    \caption{Exchange coupling and single-ion anisotropy values (in meV) used in our calculations, together with the source and the critical temperature (in K) resulting from our SRSWT.}
    \label{table:1}
   \bgroup
\def\arraystretch{1.3}
\begin{tabular*}{\linewidth}{l @{\extracolsep{\fill}} lcccccccccc}
  \hline\hline
  Material & Ref. & $J_1$ & $J_2$ & $J_3$ &$J_{z1}$ & $J_{z2}$ &$J_{z3}$ & $J_{z4}$ & $A$ & $T_\mathrm{C}^{2D}$  & $T_\mathrm{C}^{3D}$ \\
\hline
 Cr$_2$Ge$_2$Te$_6$  & \cite{gong2017n, li2018jmmm} &-2.71  &0.058  &-0.115  &0.036  &-0.086  &-0.27 &  &0.05 &27.8 &68 \\
 \hline
 \multirow{7}{1cm}{CrI$_3$}
 & \cite{chen2018prx}  &-2.01  &-0.16  &0.08  &-0.59 &   &  &  &0.22 &33.3 &51.4 \\ 
 & \cite{chen2020prb}  &-2.13 &-0.09 &0.10 &-0.59 &   & &  &0.20 & 29.7  & 48 \\
 & \cite{wang2016epl}\footnotemark[1]  &-5.5 & -1.82  & 0.2 &  &   &  &  & 0.2\footnotemark[2] &137.3 & \\ 
 & \cite{besbes2019prb}\footnotemark[1] &-1.053 &-0.373  &0.116 &-0.111 &-0.204  &-0.302 &  &0.2\footnotemark[2] &31.6  &67.4 \\ 
 & \cite{torelli20182m} &-3.24 &-0.56  &-0.001 &  &   &  &  &0.056 &56.3 & \\ 
 & \cite{zhang2015jmcc} &-2.86 &-0.64 &0.15  &  &   &  &  &0.2\footnotemark[2] &62.3 & \\ 
 & \cite{ke2021ncm}\footnotemark[1] &-2.204 &-0.356 &0.062 &-0.124 &-0.116 &-0.204 &0.222 &0.2\footnotemark[2] &44.9 &69.6 \\
 \hline\hline
\end{tabular*}
\egroup
\footnotetext[1]{Parameters from the References are rescaled, in order to account for the difference in the definitions of spin Hamiltonians.}
\footnotetext[2]{For single-ion anisotropy, missing in original paper, we take the value $0.2$ meV extracted in neutron scattering experiment~\cite{chen2020prb}.}
\end{table*}

\section{Results}

In the following, we apply the foregoing SRSWT formalism to Chromium-based layered compounds Cr$_2$Ge$_2$Te$_6$ and CrI$_3$, and find the layer-dependent magnetization vs.
temperature behavior, as well as the temperature dependence of magnon dispersion.
In these compounds, the oxidation state of Cr is $+3$, with electronic configuration [Ar]4s$^0$3d$^3$.
From Hund's rules, one can expect that Cr$^{3+}$ has a magnetic moment corresponding to the spin, $S= 3/2$.
Material-specific parameters used in our calculations, including values of exchange couplings $J_\mu$ and single-ion anisotropy $A$, are listed in Table~\ref{table:1}.
For Cr$_2$Ge$_2$Te$_6$, we employ parameters evaluated in Refs.~\cite{gong2017n, li2018jmmm} from first principles, using density functional theory (DFT).
For CrI$_3$, we use several sets of parameters reported in Refs.~\cite{chen2018prx, chen2020prb, wang2016epl, besbes2019prb, torelli20182m, zhang2015jmcc, ke2021ncm}.
These parameters are either deduced from inelastic neutron scattering experiments~\cite{chen2018prx, chen2020prb} or evaluated from first-principles, either in the framework of DFT~\cite{wang2016epl, besbes2019prb, torelli20182m, zhang2015jmcc} or using more sophisticated \textit{ab initio} methods that incorporate electron correlation effects beyond DFT~\cite{ke2021ncm}.

We obtain our results by numerically solving the self-consistency equations derived above.
The self-consistency equations contain summation over momentum.
In our simulations, we utilize a $100\times 100$ $k$-point mesh for the monolayer and few-layer systems and a $80\times 80\times 80$ mesh for the 3D system  to ensure sufficient convergence.
We check the stability of our results against the mesh size to exclude any observable finite-size effect.

\subsection{Cr$_2$Ge$_2$Te$_6$}

\begin{figure}[t]
\centerline{\includegraphics[width=\linewidth,angle=0,clip]{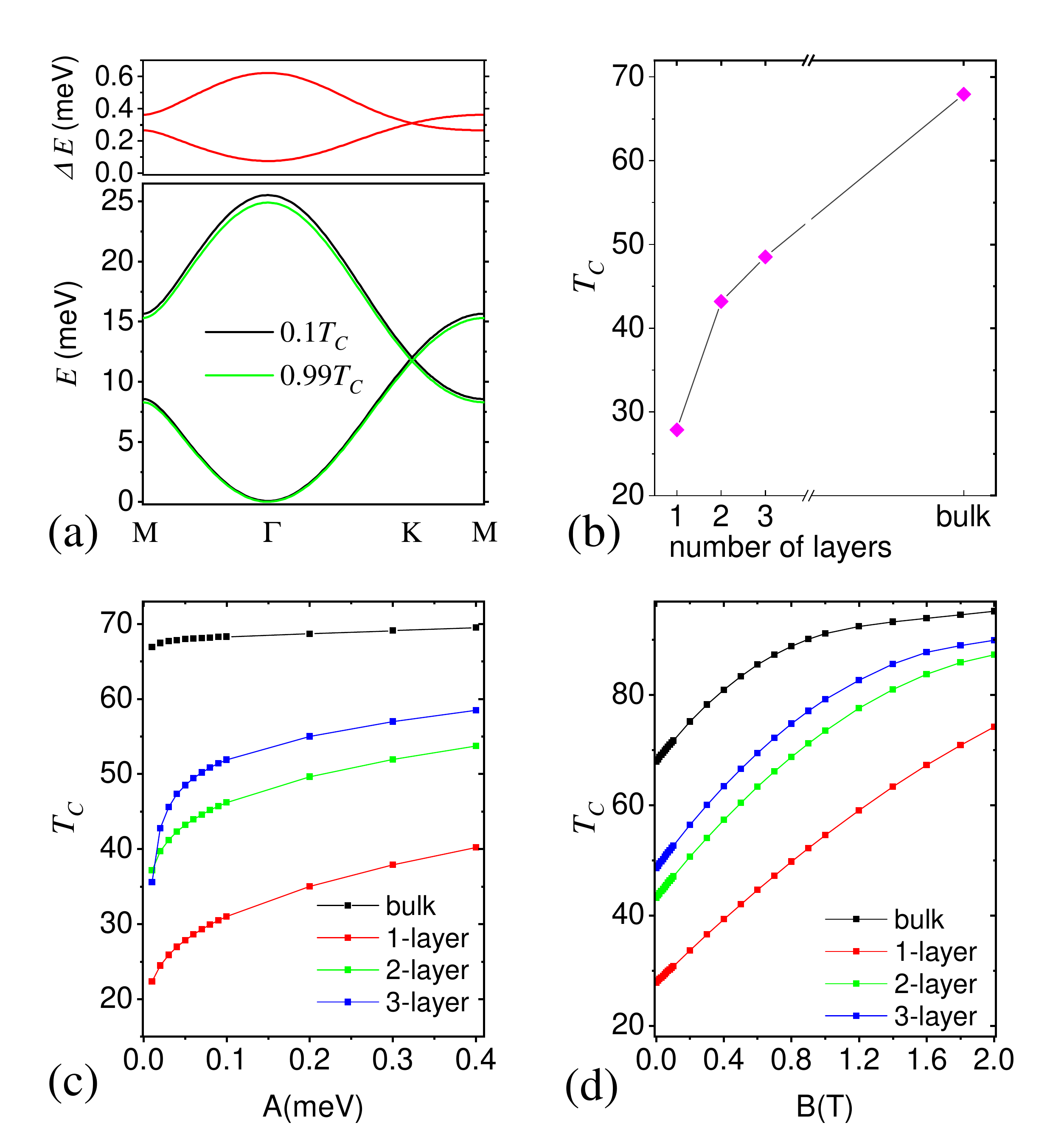}}
\caption{(Color online) SRSWT results for Cr$_2$Ge$_2$Te$_6$,
  calculated from magnetic interaction parameters of Refs.~\cite{gong2017n, li2018jmmm} (first line in Table~\ref{table:1}).
(a) Renormalized spin-wave spectrum of the monolayer at temperatures $0.1T_C$ and $0.99T_C$ is plotted with black and green lines in lower panel.
The difference of the black and green lines, $\Delta E$, is plotted in upper panel.
(b) Layer dependence of the critical temperature.
Strong dimensionality effect is observed from 2D to bulk, with critical temperature ranging from about $27.8$K to $68$K.
(c) Single-ion anisotropy dependence of the critical temperature at zero magnetic field.
In low dimensions, critical temperature sensitively depends on the value of anisotropy.
(d) Magnetic field dependence of the critical temperature, at single-ion anisotropy $A=0.05$~meV.
The non-zero value of $A$ cuts off the sharp dependence of $\tc$ on $B$ at the low-field side.}
\label{CrGeTefig}
\end{figure}

The SRSWT results for Cr$_2$Ge$_2$Te$_6$ are summarized in \rfig{CrGeTefig}.
As a hallmark of magnon self-interaction, the magnon dispersion is temperature-dependent.
This temperature dependence is illustrated in \rfig{CrGeTefig}(a).
Within the SRSWT, the temperature dependence of magnon dispersion is the direct consequence of the presence of temperature-dependent renormalization factors $(\bar{S} +f_\rho)$ in \req{Epm}.

Consistent with previous theories~\cite{corciovei1963pr, bruno1991prb, li2018jmmm}, we observe a strong dimensionality effect.
This effect is formally related to the momentum-space sums with Bose-Einstein factors [see, e.g., \req{avmag}], which are divergent in lower dimensions, unless a finite anisotropy and/or magnetic field is included.
The dimensionality effect is better seen in the dependence of critical temperature on layer number, single-ion anisotropy, and magnetic field, as shown in Figs.~\ref{CrGeTefig} (b)--(d), respectively.

Note, however, that our results are quantitatively different from those of Ref.~\cite{li2018jmmm}.
This is because in the self-consistent approach of Ref.~\cite{li2018jmmm} the short-range boson correlations $f_\rho$, \req{fjrs}, are all neglected, except for the second nearest neighbor $f_2$ (see the Discussion section for more details).

\subsection{CrI$_3$}

\begin{figure}[t]
\centerline{\includegraphics[width=\linewidth,angle=0,clip]{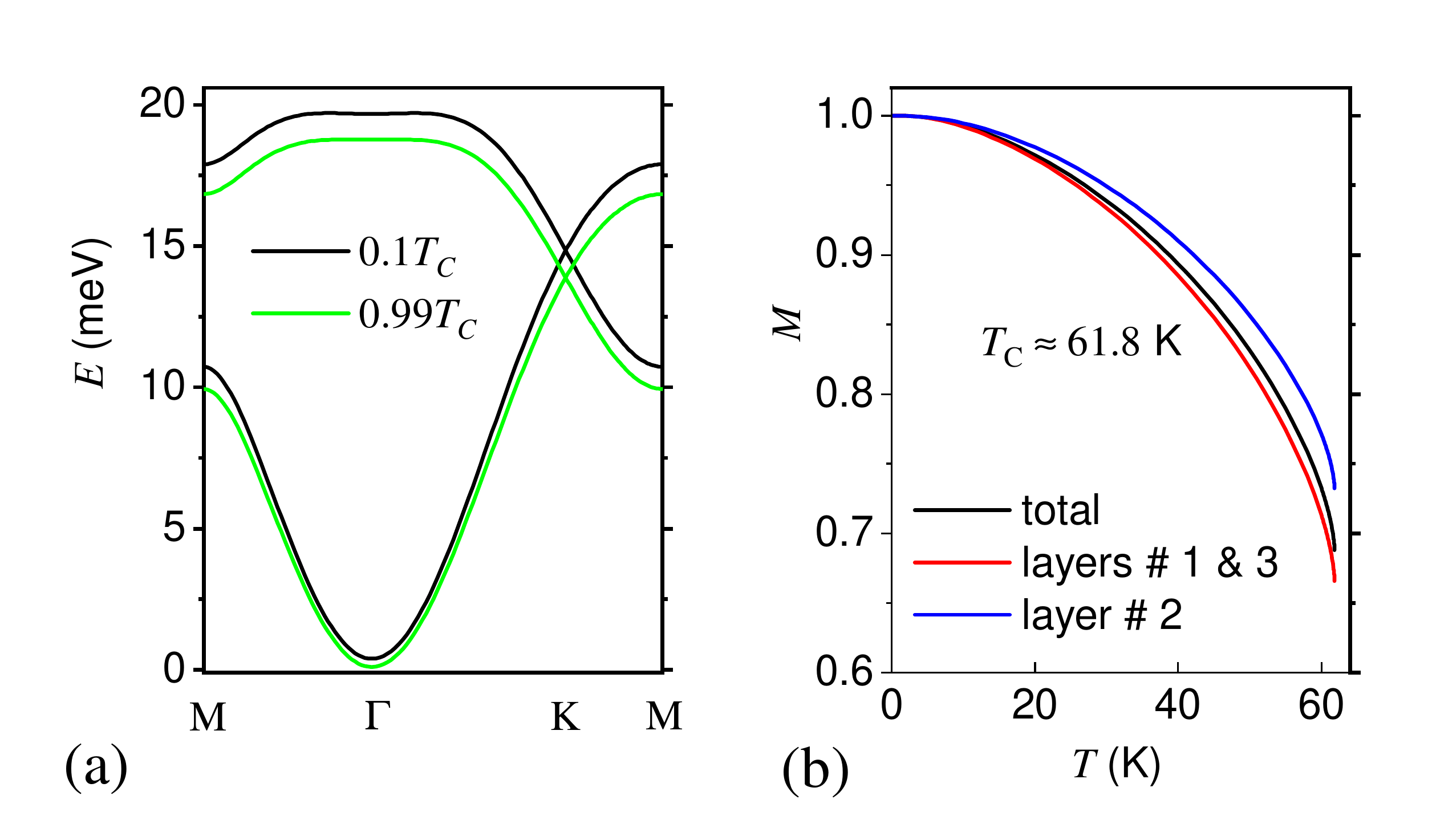}}
\caption{(Color online) SRSWT results for CrI$_3$, calculated from
magnetic interaction parameters found in Ref.~\cite{ke2021ncm} (last line in Table~\ref{table:1}).
(a) Renormalized spin-wave spectrum of the monolayer at $0.1T_C$ and $0.99T_C$.
(b) Layer-resolved temperature-dependent magnetization of the trilayer.
Magnetization on surface layers is weaker than on the bulk layer.}
\label{CrIfig}
\end{figure}

Throughout the existing literature on CrI$_3$, many reports of exchange and single-ion anisotropy values, calculated from first principles, are available.
Critical temperatures resulting from our SRSWT for these parameter values are listed in the last two columns of Table~\ref{table:1}.
These critical temperature values are quite dispersed, with some of them largely deviating from the experimentally observed ones.

For further analysis, we borrow the parameter set reported in Ref.~\cite{ke2021ncm}, which yields critical temperatures quite close to the experimental values~\cite{huang2017n}.
This parameter set includes four inter-layer exchange couplings, the strongest of which appears to be the antiferromagnetic exchange $J_{z4}$.
For further details of this unusual feature, we refer to Ref.~\cite{ke2021ncm}.

Qualitatively, our results for CrI$_3$ are quite similar to the ones for Cr$_2$Ge$_2$Te$_6$ in \rfig{CrGeTefig}.
We observe a strong dimensionality effect, with the critical temperature growing from about 45 K to 70 K as one goes from monolayer to bulk.
As in the previous case, critical temperatures sensitively depend on the external magnetic field and single-ion anisotropy in lower dimensions.
One of the distinctive features that we encounter for CrI$_3$ is the temperature dependence of the renormalization of magnon spectrum of the monolayer, shown \rfig{CrIfig}(a), which is notably stronger than that of Cr$_2$Ge$_2$Te$_6$.

Another remarkable difference from the previous case is that the layer-resolved magnetization shows a larger deviation of magnetization between the bulk and surface layers.
This is illustrated in \rfig{CrIfig}(b), where the magnetization of the trilayer CrI$_3$ is plotted against temperature.
The weaker magnetization of surface layers in \rfig{CrIfig}(b) is what we typically see in our SRSWT simulations for few-layer systems with various number of layers, for both materials considered.
This dependence is consistent with the long-known results on magnetic thin films~\cite{wolfram1972pss} and can be traced back to the higher magnon density on surface layers as compared to the bulk.

\section{Discussion}

Magnetic properties of Cr$_2$Ge$_2$Te$_6$ have been recently investigated by Li {\it et al.}~\cite{li2018jmmm}, using a self-consistent mean-field scheme (see also Ref.
\cite{gong2017n}).
The approach adopted in this work is different from the theory of Refs.~\cite{gong2017n, li2018jmmm} in two ways; 1) The Hartree-Fock-like decoupling of quartic terms in Refs.
\cite{gong2017n, li2018jmmm} is performed by keeping terms diagonal in {\em both} momentum and sublattice spaces.
Thus, all inter-sublattice correlations, which are diagonal in momentum but non-diagonal in sublattice space, are ignored.
This implies that all short-range correlations $f^\rho_{\nu \nu'}$ with $\nu\neq\nu'$ are set to zero.
Here we keep terms diagonal in momentum space, including those non-diagonal in sublattice indices.
As a result, our self-consistency equations contain at least one thermodynamic quantity $f^\rho_{\nu \nu'}$ per each exchange coupling $J_\rho$.
2) In Refs.~\cite{gong2017n, li2018jmmm}, theoretical analysis of few-layer systems with layer stacking along the $z$ direction is based upon a Fourier transform in the $z$ direction.
Thus, it is assumed that the system is periodic in that direction.
However, this conflicts with the very nature of few-layer systems where the surface layers are physically different from the inner layers.
In contrast, we treat the layers as separate sublattices.
This approach allows us to account for the physical difference between surface and inner layers.

In the model under consideration, we incorporate a uniaxial single-ion anisotropy.
Anisotropic interaction is crucial in lower dimensions, where it opens up a spin-wave gap rendering the magnetic ordering possible at non-zero temperatures.
The single-ion anisotropy term in \req{coeffsAB} and its 3D and quasi-2D counterparts is $\propto(4\bar{S} -2S -1)$.
One drawback related to this term is that it may turn to zero at a temperature lower than the true ordering temperature, resulting in a spuriously vanishing spectral gap and destroying the theory'validity near $\tc$.
This issue is typical to renormalized spin-wave theories~\cite{anderson1964pr, rastelli1974jpcs, pini1981jpcs, irkhin1999prb} and, besides the error coming from the Hartree-Fock approximation, it is related to the violation of kinematical restrictions while treating the bosonic excitations as independent bosons and including unphysical states with a high number of bosons.

One way of resolving this issue is by replacing the factor $(4\bar{S} -2S -1)$ with other forms, based on some physical arguments.
For example, a form of the single-ion anisotropy, corresponding to the replacement
\begin{equation}
\label{ACreplace}
 (4\bar{S} -2S -1) \to \left(2\bar{S} +(\bar{S}/S)^2(2\bar{S}-2S -1)\right),
\end{equation}
emerges due to the Anderson -- Callen decoupling~\cite{anderson1964pr}, commonly used in the Green's function approach to the Heisenberg model~\cite{ froebrich2006prep, jensen2006ssr}.
Unlike the left-hand side of \req{ACreplace}, its right-hand side vanishes only at $\bar{S}=0$.
In support of the above replacement is also the fact that the right-hand side of \req{ACreplace} converges to the left-hand side upon the large-$S$ expansion.

Another consequence of the violation of kinematical restrictions is that the single-ion anisotropy does not vanish for $S=1/2$, as it should.
In an effort to remedy this issue, the replacement
\begin{equation}
\label{IKKreplace}
 (4\bar{S} -2S -1) \to (2S-1)\left(\bar{S}/S\right)^2
\end{equation}
was suggested in Ref.~\cite{irkhin1999prb}, with the justification that the right-hand side of \req{IKKreplace} contains the necessary factor, $(2S-1)$, and is equivalent to the left-hand side within the large-$S$ expansion.

In the parametric domain considered in this work for the two chromium-based materials, the factor $(4\bar{S} -2S -1)$ does not turn to zero at a temperature lower than $\tc$.
Nevertheless, we have performed calculations using replacements Eqs.
(\ref{ACreplace}) and (\ref{IKKreplace}) in \req{coeffsAB} and its 3D and quasi-2D counterparts.
As expected, these replacements have very little effect on the magnetization curve $M(T)$ at the lowest temperatures.
Still, they induce an appreciable increase of the critical temperature by about 10 to 15\% in a zero magnetic field.

The SRSWT presented above corresponds to the summation of all bubble graphs to approximate the one-particle boson Green function~\cite{ loly1971jpcs}.
This approximation does not capture the interaction-induced magnon lifetime.
Although it is beyond the scope of the present work, here we sketch a direct way to calculate the interaction-induced magnon lifetime \cite{liu1992jpcm, costafilho2000prb}.
To the leading order, the magnon lifetime may be found from the spin-wave interaction Hamiltonian $H_4$, \req{H1}, by going beyond the Hartree-Fock approximation.
This can be done by representing the interacting spin-wave Hamiltonian, $\bar{H}=H_2+H_4$, as
\begin{equation}
\label{Htot}
 \bar{H} = H_R+V,
\end{equation}
where $H_R$ is the mean-field part (the renormalized Hamiltonian), \req{HMF}, and $V=H_2+H_4-H_R$ is the interaction part beyond the mean-field.
Furthermore, $V$ can be represented as the interaction between the renormalized magnons -- eigenmodes of $H_R$, and the corresponding interaction corrections can be found by calculating the renormalized magnon self-energy.


\section{Conclusion}

We developed a self-consistently renormalized spin-wave theory for the ferromagnetic Heisenberg model with perpendicular easy-axis single-ion anisotropy, defined on monolayer, few-layer, and bulk systems with honeycomb in-plane arrangement of spins.
We treat the layers of a few-layer system as sublattices.
This approach allows us to account for the difference of surface and bulk layers and pinpoint the different strengths of average magnetization on the surface and bulk layers.
In addition, our approach can be directly generalized to study systems with magnetically inequivalent sublattices.

We have applied the developed theory to Chromium-based layered ferromagnetic compounds Cr$_2$Ge$_2$Te$_6$ and CrI$_3$, for which experimental results are available for bulk and mechanically exfoliated few-layer samples (down to the monolayer in the case of CrI$_3$).
We have calculated the magnetization dependence on the temperature, $M(T)$, and the Curie temperature $\tc$ for these materials.
Our calculations have used sets of reported magnetic interaction values for the two materials obtained from first principles, and for CrI$_3$, also from neutron scattering experiments.
For different sets of exchange parameters, we find quite different values of $\tc$, both for monolayer and bulk configurations, also deviating from the experimental  $\tc$ values.
Despite the quantitative discrepancy, we encounter a strong dimensionality effect with the critical temperature sensitively depending on the number of layers, and enhanced sensitivity to the magnetic field and single-ion anisotropy strength in lower dimensions, consistent with experimental observations.

\section*{Acknowledgments}
This work was supported by the U.S.~Department of Energy, Office of Science, Office of Basic Energy Sciences, Materials Sciences and Engineering Division, and Early Career Research Program.
Ames Laboratory is operated for the U.S.~Department of Energy by Iowa State University under Contract No.~DE-AC02-07CH11358.

\appendix

\section{Details of derivations}

In this Appendix we present basic details of our derivations and discuss specific features of the theory for the monolayer, bulk, and few-layer (quasi-2D) systems.

The derivation of \req{HMF} from either of the Holstein-Primakoff or the Dyson-Maleev transformation followed by the Hartree-Fock decoupling \req{HF4bt} is quite straightforward, so we skip this part and start with the derivation of \req{univcoeffs}.

Consider the sum, $N^{-1}\sum_{\mathbf{k}' } J^{\mathbf{k} -\mathbf{k}'}_{\nu \nu'} n_{\mathbf{k}' \nu' \nu}$, which appears in the first line of \req{coeffs}, as well as in the last line of the same equation, with $\mathbf{k}=0$.
Using Eqs.~(\ref{univgamma}) and (\ref{univJk}), this sum may be cast in the form,
\begin{equation}
\label{sumJnk}
  \sum_{\rho \in [\nu \nu']}J_\rho\sum_{\mathbf{u}^\rho_{\nu \nu'}}
  e^{i\mathbf{k} \cdot \mathbf{u}^\rho_{\nu \nu'}}\left( \frac 1N\sum_{\mathbf{k}'
  }e^{-i\mathbf{k}' \cdot \mathbf{u}^\rho_{\nu \nu'}} n_{\mathbf{k}' \nu' \nu} \right).
\end{equation}
The sum over $\mathbf{k}'$ in \req{sumJnk} is independent of
$\mathbf{u}^\rho_{\nu \nu'}$ by symmetry, so the term in the
parentheses is equivalent to the right-hand side of \req{univfis}. Using \req{univgamma} for the remaining
terms of \req{sumJnk}, one gets
\begin{equation}
\label{Jnkwf}
  \frac 1N\sum_{\mathbf{k}' } J^{\mathbf{k} -\mathbf{k}'}_{\nu \nu'} n_{\mathbf{k}'
  \nu' \nu}= \sum_{\rho \in [\nu \nu']}\gamma^\rho_{\nu
  \nu'}(\mathbf{k})f^\rho_{\nu \nu'}.
\end{equation}
Utilizing Eqs.~(\ref{univgamma})--(\ref{slavmag}) and (\ref{Jnkwf}) in \req{coeffs}, one encounters the relation \req{univcoeffs}.
Furthermore, alternatively to the steps succeeding \req{univcoeffs}, the self-consistency could be formulated as the condition of saddle point for
\begin{eqnarray}
\label{fnctnl}
  \Omega=\frac1{\beta N} \sum_{\mathbf{k}, \sigma}
  \ln\left(1-e^{-\beta E_{\sigma}(\mathbf{k})} \right)
  +2A\sum_\nu\left(\bar{S}_\nu -S\right)^2\quad && \\
  -\! \sum_{\nu, \nu'}\sum_{\rho \in [\nu\nu']}\!
  \frac{\gamma^\rho_{\nu \nu'}(0)}2\!\left(\bar{S}_\nu +f^\rho_{\nu \nu'}
  -S\right)\! \left(\bar{S}_{\nu'} +f^\rho_{\nu' \nu} -S\right)\! ,&& \nonumber
\end{eqnarray}
which is closely related to the Helmholtz free energy of the system per unit cell.

Further details of the SRSWT for the three different configurations discussed in the main text follow from the specific structure factors which in turn are determined by the non-zero exchange couplings.

\subsection{Monolayer}

\label{App2D}

For the monolayer, we distinguish three non-zero intra-plane exchange couplings $J_1$, $J_2$, and $J_3$, indicated in \rfig{fewlayers}, and two sublattices labeld by $\nu, \nu' =1, 2$.
The corresponding linking vectors $\mathbf{u}^\rho_{\nu \nu'}$ are
\begin{eqnarray}
\label{2Dus}
  && \{\mathbf{u}_{1}\} \equiv \{\mathbf{u}^1_{1 2}\} = \boldsymbol{\delta}_1,\,
  \boldsymbol{\delta}_2,\, \boldsymbol{\delta}_3, \nonumber \\
  && \{\mathbf{u}_{2}\} \equiv \{\mathbf{u}^2_{\nu \nu}\} =
  \pm \mathbf{a}_1,\, \pm \mathbf{a}_2,\, \pm (\mathbf{a}_1 - \mathbf{a}_2), \nonumber \\
  && \{\mathbf{u}_{3}\} \equiv \{\mathbf{u}^3_{1 2}\}= -2\boldsymbol{\delta}_1,\,
  -2\boldsymbol{\delta}_2,\, -2\boldsymbol{\delta}_3,
\end{eqnarray}
with $\boldsymbol{\delta}_i$ and $\mathbf{a}_i$ shown in \rfig{fewlayers}.
Vectors \req{2Dus} lead to the structure factors and thermodynamic quantities given by Eqs.
(\ref{2Dgammas}), (\ref{2DSf}).

\subsection{Bulk}

%
\label{App3D}
The bulk system considered here involves three intra-layer, $J_1$, $J_2$, $J_3$, and up to four inter-layer exchange couplings, $J_{z1}$, $J_{z2}$, $J_{z3}$, and $J_{z4}$, specified in \rfig{fewlayers}.
The crystal structure is spanned by the sublattice primitive vectors,
\begin{eqnarray}
\label{ais}
  &&\mathbf{a}_1=  \frac a2\left(3, \sqrt{3},0 \right), \nonumber
  \\
  &&\mathbf{a}_2= \frac a2\left(3, -\sqrt{3},0 \right), \nonumber
  \\
  &&\mathbf{a}_3= a\left(1, 0, \frac ca \right),
\end{eqnarray}
where $a$ is the intra-layer magnetic atom separation and $c$ is the layer spacing.
The subsequent analysis is customarily based on the introduction of the reciprocal momentum space.
Note however that the 3D Fourier transform to the momentum space implies periodic boundary conditions in all, including the out-of-plane, directions.

The bulk system is readily described by Eqs.~(\ref{HMF}) and (\ref{univcoeffs}), with properly specified structure factors and thermodynamic quantities.
Importantly, the system is still comprising of only two equivalent sublattices.
This leaves us with the total of seven structure factors (one per each exchange coupling) including the three intra-layer structure factors given by \req{2Dgammas}, and four inter-layer ones,
\begin{equation}
\label{gammazs}
  \gamma_\rho(\mathbf{k})= J_\rho\sum_ {\mathbf{u}_\rho}
  e^{i\mathbf{k} \cdot \mathbf{u}_\rho},\quad \rho=z1,\,z2,\,z3,\,z4,
\end{equation}
with $\mathbf{u}_\rho$ running over the inter-layer links coupled
by the exchange $J_\rho$,
\begin{eqnarray}
\label{3Dus}
  &&\mathbf{u}_{z1}\equiv \mathbf{u}^{z1}_{1 2} = \mathbf{a}_3 + \boldsymbol{\delta}_3, \nonumber\\
  && \{ \mathbf{u}_{z2} \}\equiv \{\mathbf{u}^{z2}_{\nu \nu}\} =
  \pm \mathbf{a}_3,\, \pm (\mathbf{a}_3 - \mathbf{a}_1),\, \pm (\mathbf{a}_3 -
  \mathbf{a}_2), \nonumber \\
  &&\{ \mathbf{u}_{z3} \}\equiv \{\mathbf{u}^{z3}_{1 2}\} = (\boldsymbol{\delta}_1 -\mathbf{a}_3),\,
  (\boldsymbol{\delta}_2-\mathbf{a}_3),\, -(\mathbf{a}_3
  +2\boldsymbol{\delta}_3),\quad \nonumber \\
  &&\{ \mathbf{u}_{z4} \}\equiv \{\mathbf{u}^{z4}_{1 2}\} = (\mathbf{a}_3 +\boldsymbol{\delta}_1),\,
  (\mathbf{a}_3 -\mathbf{a}_1 + \boldsymbol{\delta}_2), \nonumber \\
  &&\hspace{2.7cm} (\mathbf{a}_3 -\mathbf{a}_2 +\boldsymbol{\delta}_3).
\end{eqnarray}
Due to the equivalence of the two sublattices, the average magnetization is expected to be the same on both sublattices; see the first line of \req{2DSf}.
The remaining seven thermodynamic quantities are the three intra-layer $f_i$, $i=1$, $2$, $3$, given in \req{2DSf}, and four more, inter-layer ones,
\begin{equation}
\label{3Dfs}
 f_{z1}\equiv f^{z1}_{1 2}, \quad f_{z2}\equiv f^{z2}_{\nu \nu},\quad
 f_{z3}\equiv f^{z3}_{1 2}, \quad f_{z4}\equiv f^{z4}_{1 2}.
\end{equation}
Furthermore, the renormalized Hamiltonian of the bulk system is
given by \req{H2D}, with
\begin{eqnarray}
\label{3DcoeffsAB}
  \mathcal{A}({\mathbf{k}}) &&= g\mu_B B + \left( \bar{S} +f_2
  \right)\gamma_2(\mathbf{k})+ \left( \bar{S} +f_{z2}
  \right)\gamma_{z2}(\mathbf{k})\nonumber \\
   &&- \sum_\rho\left( \bar{S} + f_\rho\right)\gamma_\rho(0)
   -\left (2S+1 -4\bar{S} \right)A,   \nonumber \\
  \mathcal{B}({\mathbf{k}}) &&= \sum_\mu\left (\bar{S} +f_\mu \right
  )\gamma_\mu(\mathbf{k}),
\end{eqnarray}
where the index $\rho$ runs over the seven values $1$, $2$, $3$, $z1$, $z2$, $z3$, $z4$, and $\mu$ -- over the five inter-sublattice values, $1$, $3$, $z1$, $z3$, $z4$.
Similar to that in 2D, the magnon dispersion is
\begin{equation}
\label{Ek3D}
  E_{\pm}(\mathbf{k})= \mathcal{A}({\mathbf{k}}) \pm
  |\mathcal{B}({\mathbf{k}})|,
\end{equation}
the average magnetization is given by \req{avmag}, and
$f_\mu$ are given by
\begin{equation}
\label{fifzi3D0}
  f_\mu =\frac 1{2N \gamma_\mu(0)}\sum_{\sigma=\pm}\sum_{\mathbf{k}}
  \frac {\gamma_\mu(\mathbf{k})}{\exp\left[\beta
  E_{\sigma}(\mathbf{k})\right]-1},
\end{equation}
for $\mu=2$, $z2$ (intra-sublattice $f$'s) and
\begin{equation}
\label{fifzi3D}
  f_\mu = \frac 1{2N \gamma_\mu(0)}\sum_{\sigma=\pm}\sum_{\mathbf{k}}
  \frac {\sigma\gamma_\mu(\mathbf{k}) e^{-i\phi_\mathbf{k}}}{\exp\left[\beta
  E_{\sigma}(\mathbf{k})\right]-1},
\end{equation}
for $\mu=1$, $3$, $z1$, $z3$, $z4$ (inter-sublattice $f$'s), where
\begin{equation}
\label{3Dfk}
  \phi_\mathbf{k}=\arg \left[ \mathcal{B} ({\mathbf{k}}) \right].
\end{equation}
Equations (\ref{avmag}), (\ref{3DcoeffsAB}) -- (\ref{3Dfk}) form a closed set of self-consistency equations from which the average magnetization $\bar{S}$ is found for the bulk 3D system, at a given temperature.

\subsection{The few-layer system}

%
\label{AppQ2D}
The $L$ -- layer system is treated as a system of $2L$ sublattices.
The matrix elements of $\hat{H}_\mathbf{k}$, \req{Hkmrx}, may be read off of \req{univcoeffs}.
The diagonal $\hat{h}$-operators are of the form,
\begin{equation}
\label{hathq2D}
  \hat{h}_l= \! \left ( \! \!
  \begin{array}{cc}
  \mathcal{A}_{2l-1}(\mathbf{k}) &\mathcal{B}_l(\mathbf{k})\\
  \mathcal{B}_l^*(\mathbf{k}) &\mathcal{A}_{2l}(\mathbf{k})
  \end{array} \! \! \right )\! ,
\end{equation}
where
\begin{eqnarray}
\label{Aodd}
  \mathcal{A}_{2l-1}({\mathbf{k}}) &&=g\mu_B B -\left (2S+1 -4\bar{S}_{2l-1} \right)A  \nonumber \\
  &&- \left( \bar{S}_{2l-1} +f^2_{(2l-1)(2l-1)} \right) \bigl(\gamma_2(0)
  -\gamma_2(\mathbf{k})\bigr)\quad
    \nonumber \\
  &&- \left( \bar{S}_{2l} +\text{Re}\bigl[f^1_{(2l-1)(2l)} \bigr] \right)
  \gamma_1(0)\nonumber \\
  &&- \left( \bar{S}_{2l} +\text{Re}\bigl[f^3_{(2l-1)(2l)} \bigr] \right)
  \gamma_3(0)\nonumber \\
  &&- \left( \bar{S}_{2l+2} +\text{Re}\bigl[f^{z1}_{(2l-1)(2l+2)} \bigr] \right)
  \gamma_{z1}(0)\nonumber \\
  &&- \left( \bar{S}_{2l+1} +\text{Re}\bigl[f^{z2}_{(2l-1)(2l+1)} \bigr] \right)
  \gamma_{z2}(0)\nonumber \\
  &&- \left( \bar{S}_{2l-3} +\text{Re}\bigl[f^{z2}_{(2l-1)(2l-3)} \bigr] \right)
  \gamma_{z2}(0)\nonumber \\
  &&- \left( \bar{S}_{2l-2} +\text{Re}\bigl[f^{z3}_{(2l-1)(2l-2)} \bigr] \right)
  \gamma_{z3}(0)\nonumber \\
  &&- \left( \bar{S}_{2l+2} +\text{Re}\bigl[f^{z4}_{(2l-1)(2l+2)} \bigr] \right)
  \gamma_{z4}(0), \\
  \label{Aeven}
  \mathcal{A}_{2l}({\mathbf{k}}) &&= g\mu_B B -\left (2S+1 -4\bar{S}_{2l} \right)A  \nonumber \\
  &&- \left( \bar{S}_{2l} +f^2_{(2l)(2l)} \right) \bigl(\gamma_2(0) -\gamma_2(\mathbf{k})\bigr)
    \nonumber \\
  &&- \left( \bar{S}_{2l-1} +\text{Re}\bigl[f^1_{(2l)(2l-1)} \bigr] \right)
  \gamma_1(0)\nonumber \\
  &&- \left( \bar{S}_{2l-1} +\text{Re}\bigl[f^3_{(2l)(2l-1)} \bigr] \right)
  \gamma_3(0)\nonumber \\
  &&- \left( \bar{S}_{2l-3} +\text{Re}\bigl[f^{z1}_{(2l)(2l-3)} \bigr] \right)
  \gamma_{z1}(0)\nonumber \\
  &&- \left( \bar{S}_{2l+2} +\text{Re}\bigl[f^{z2}_{(2l)(2l+2)} \bigr] \right)
  \gamma_{z2}(0)\nonumber \\
  &&- \left( \bar{S}_{2l-2} +\text{Re}\bigl[f^{z2}_{(2l)(2l-2)} \bigr] \right)
  \gamma_{z2}(0)\nonumber \\
  &&- \left( \bar{S}_{2l+1} +\text{Re}\bigl[f^{z3}_{(2l)(2l+1)} \bigr] \right)
  \gamma_{z3}(0)\nonumber \\
  &&- \left( \bar{S}_{2l-3} +\text{Re}\bigl[f^{z4}_{(2l)(2l-3)} \bigr] \right)
  \gamma_{z4}(0),
\end{eqnarray}
for the bulk layers with $l=2,\cdots,2L-1$, whereas for the surface layers ($l=1$ and $2L$) the lines containing subscripts less than $1$ or greater than $2L$ in Eqs.
(\ref{Aodd}), (\ref{Aeven}) are omitted.
The off-diagonals of \req{hathq2D} are given by
\begin{eqnarray}
  \label{Bl}
  \mathcal{B}_l({\mathbf{k}}) &&= \left( \frac {\bar{S}_{2l-1} + \bar{S}_{2l}}2
  +f^1_{(2l-1)(2l)}\right) \gamma_1(\mathbf{k})\nonumber \\
  &&+ \left( \frac {\bar{S}_{2l-1} + \bar{S}_{2l}}2
  +f^3_{(2l-1)(2l)}\right) \gamma_3(\mathbf{k}).\quad
\end{eqnarray}
The off-diagonal elements of $\hat{H}_\mathbf{k}$ describing the inter-layer couplings have the matrix form,
\begin{equation}
\label{hatb}
  \hat{b}_l= \left (
  \begin{array}{cc}
  \mathcal{B}^1_l(\mathbf{k}) &\mathcal{B}^2_l(\mathbf{k})\\
  \mathcal{B}^3_l(\mathbf{k}) &\mathcal{B}^4_l(\mathbf{k})
  \end{array} \right ),\quad l=1,\cdots, L-1,
\end{equation}
with
\begin{eqnarray}
  \label{Bil}
  \mathcal{B}_l^1({\mathbf{k}}) &&= \left( \frac {\bar{S}_{2l-1} + \bar{S}_{2l+1}}2
  +f^{z2}_{(2l-1)(2l+1)}\right) \gamma_{z2}(\mathbf{k}),\nonumber \\
  \mathcal{B}_l^2({\mathbf{k}}) &&= \left( \frac {\bar{S}_{2l-1} + \bar{S}_{2l+2}}2
  +f^{z1}_{(2l-1)(2l+2)}\right) \gamma_{z1}(\mathbf{k})\nonumber \\
  &&+ \left( \frac {\bar{S}_{2l-1} + \bar{S}_{2l+2}}2
  +f^{z4}_{(2l-1)(2l+2)}\right) \gamma_{z4}(\mathbf{k}),\nonumber \\
  \mathcal{B}_l^3({\mathbf{k}}) &&= \left( \frac {\bar{S}_{2l} +
  \bar{S}_{2l+1}}2 +f^{z3}_{(2l)(2l+1)}\right) \gamma_{z3}(\mathbf{k}),\nonumber \\
  \mathcal{B}_l^4({\mathbf{k}}) &&= \left( \frac {\bar{S}_{2l} + \bar{S}_{2l+2}}2
  +f^{z2}_{(2l)(2l+2)}\right) \gamma_{z2}(\mathbf{k}).
\end{eqnarray}

By noting that, similarly to Eqs.~(\ref{2Dgammas}) and (\ref{gammazs}), $\gamma^\rho_{\nu \nu'}(\mathbf{k})$ with $\nu\leq\nu'$ is completely specified by the single index $\rho$, in Eqs.
(\ref{Aodd})--(\ref{Bil}) we have only seven structure factors, as before.
The intra-layer structure factors $\gamma_1(\mathbf{k})$, $\gamma_2(\mathbf{k})$, and $\gamma_3(\mathbf{k})$ are the same as in the two previous cases of the monolayer and bulk systems, given by \req{2Dgammas}.
However, the inter-layer structure factors are somewhat different from those for the bulk 3D case, \req{gammazs}, because the layers are treated as separate 2D sublattices.
We have
\begin{equation}
\label{gammazsq2D}
  \gamma_\rho(\mathbf{k})= J_\rho\sum_ {\mathbf{v}_\rho}
  e^{i\mathbf{k}\cdot \mathbf{v}_\rho},\quad \rho=z1,\, z2,\, z3,\, z4,
\end{equation}
where the notation $\mathbf{v}_\rho$ is used instead of the more general $\mathbf{u}^\rho_{\nu \nu'}$ for the vectors running over the inter-sublattice links coupled by the exchange $J_\rho$.
The relation between $\mathbf{v}_\rho$ and $\mathbf{u}^\rho_{\nu \nu'}$ and their explicit forms are
\begin{eqnarray}
\label{vziq2D}
  &&\mathbf{v}_{z1} \equiv\mathbf{u}^{z1}_{(2l-1)(2l+2)}  = \boldsymbol{\delta}_3, \nonumber \\
  && \{ \mathbf{v}_{z2} \} \equiv \{
  \mathbf{u}^{z2}_{(2l-1)(2l+1)}\}\equiv \{ \mathbf{u}^{z2}_{(2l)(2l+2)}\}
  = 0,\,  - \mathbf{a}_1,\, - \mathbf{a}_2, \nonumber \\
  &&\{ \mathbf{v}_{z3} \} \equiv \{ \mathbf{u}^{z3}_{(2l)(2l+1)}\}
  = - \boldsymbol{\delta}_1,\, - \boldsymbol{\delta}_2,\,
  2\boldsymbol{\delta}_3, \nonumber \\
  &&\{ \mathbf{v}_{z4} \}\equiv \{ \mathbf{u}^{z4}_{(2l-1)(2l+2)}\}
  = \boldsymbol{\delta}_1,\,
  (-\mathbf{a}_1 + \boldsymbol{\delta}_2),\, (-\mathbf{a}_2
  +\boldsymbol{\delta}_3),\nonumber
\end{eqnarray}
independently of $l$, for $l=1,\cdots, L-1$.
The short-range correlators in Eqs.
(\ref{Aodd})--(\ref{Bil}) are defined by \req{univfis}, and the self-consistency is established by solving the eigenvalue problem \req{numdiag} and utilizing $\Lambda_\mathbf{k}$ in Eqs.
(\ref{scSs}) and (\ref{scfs}).
Thus, the self-consistency relates a total of $11L-5$ unknown thermodynamic quantities: $2L$ sublattice magnetizations $\bar{S}_\nu$ and $9L-5$ short-range correlations $f^\rho_{\nu \nu'}$, occurring in Eqs.
(\ref{Aodd})--(\ref{Bil}).
Furthermore, as mentioned in the text, sublattices with subscripts $\nu$ and $2L+1-\nu$ are eqivalent, reducing the total number of independent variables to $C_L=(11L-3)/2$ or $C_L=(11L-2)/2$ for odd or even $L$, respectively.

\bibliography{aaa}
\bigskip

\end{document}